\documentclass[preprint,5p,times,twocolumn]{elsarticle}
\usepackage{graphicx} 
\usepackage{epstopdf} 
\usepackage{dcolumn} 
\usepackage{xcolor} 
\usepackage{amsmath,amssymb} 
\usepackage{hyperref} 
\usepackage[utf8]{inputenc} 
\usepackage[english]{babel} 
\usepackage[displaymath, mathlines]{lineno} 
\usepackage{blindtext} 
\usepackage{ifthen} 
\usepackage{orcidlink} 

\newboolean{articletitles}
\setboolean{articletitles}{true} 

\input{belle2-symbols}

\journal{NIM-A}

\begin{document}

\title{Kink Finder at Belle II
}

\author{D.~Bodrov\,\orcidlink{0000-0001-5279-4787}} 
\author{X.P.~Xu\,\orcidlink{0000-0001-5096-1182}} %
\author{D.~Gavrilov\,\orcidlink{0000-0003-3249-5813}} 
\author{P.~Pakhlov\,\orcidlink{0000-0001-7426-4824}} 
\author{V.~Bertacchi\,\orcidlink{0000-0001-9971-1176}}
\author{T.~Bilka\,\orcidlink{0000-0003-1449-6986}}
\author{A.~Biswas\,\orcidlink{0009-0002-6336-5640}}
\author{G.~Casarosa\,\orcidlink{0000-0003-4137-938X}}
\author{P.~Cheema\,\orcidlink{0000-0001-8472-5727}}
\author{L.~Corona\,\orcidlink{0000-0002-2577-9909}}
\author{G.~De~Pietro\,\orcidlink{0000-0001-8442-107X}}
\author{T.~V.~Dong\,\orcidlink{0000-0003-3043-1939}}
\author{P.~Ecker\,\orcidlink{0000-0002-6817-6868}}
\author{T.~Ferber\,\orcidlink{0000-0002-6849-0427}}
\author{R.~Godang\,\orcidlink{0000-0002-8317-0579}}
\author{A.~Heidelbach\,\orcidlink{0000-0002-6663-5469}}
\author{T.~Lam\,\orcidlink{0000-0001-9128-6806}}
\author{M.~Laurenza\,\orcidlink{0000-0002-7400-6013}}
\author{T.~Lueck\,\orcidlink{0000-0003-3915-2506}}
\author{L.~Massaccesi\,\orcidlink{0000-0003-1762-4699}}
\author{F.~Meier\,\orcidlink{0000-0002-6088-0412}}
\author{L.~Reuter\,\orcidlink{0000-0002-5930-6237}}
\author{B.~Scavino\,\orcidlink{0000-0003-1771-9161}}
\author{K.~Schoenning\,\orcidlink{0000-0002-3490-9584}}
\author{J.~Skorupa\,\orcidlink{0000-0002-8566-621X}}
\author{S.~Spataro\,\orcidlink{0000-0001-9601-405X}}
\author{T.~Tien-Manh\,\orcidlink{0009-0002-6463-4902}}
\author{V.~Vobbilisetti\,\orcidlink{0000-0002-4399-5082}}
\author{C.~Wessel\,\orcidlink{0000-0003-0959-4784}}
\author{J.~Wiechczynski\,\orcidlink{0000-0002-3151-6072}}
\begin{abstract}
We present a track-finding algorithm for the Belle II experiment that specifically targets so-called kinks: signatures of charged particles decaying or scattering in-flight in the detector material, resulting in a sudden and significant change of the particle's flight direction. Our benchmark studies of this \textit{Kink Finder} show that the reconstruction efficiency for such signatures is about 40\%, compared to a value of around 11\% for the standard Belle II track-finding algorithm. Our studies also show that the Kink Finder significantly improves the resolution of the secondary track parameters, suppresses the number of cloned tracks, and reduces the PID misidentification rates for kaon and pions.
\end{abstract}

\begin{keyword}
Belle~II, tracking, in-flight decays, kinks
\end{keyword}

\maketitle

\section{Introduction}
\label{sec:intro}

The Belle~II experiment is conducted at the SuperKEKB accelerator in Tsukuba, Japan. SuperKEKB is a high-luminosity asymmetric energy $e^+e^-$ collider with a 4 GeV positron beam and a 7 GeV electron beam, providing a center-of-mass energy of around 10.58~GeV. The Belle~II experiment enables precision studies of flavor physics and rare decays and direct searches for physics beyond the Standard Model (e.g., forbidden decays, dark sector particles), including through the application of novel methods. These goals require the development of high-quality hardware and software. 

The reconstruction of a so-called kink -- the signature of a particle that decays or scatters in-flight, leading to a sudden and significant change in the direction of motion -- is one of the steps to improve the performance of the experiment. Furthermore, it increases the scope of the experiment, enabling the measurement of hitherto unmeasured physics processes.
Processes that result in kinks include \kmunu\footnote{Charge conjugation is implied throughout the paper unless otherwise indicated.}, \kpipi, \kpipipi, \kcpi, \kpmunu, \kpenu, \pimunu, and \menn, but also scattering in the detector material. Kink reconstruction can reduce the misidentification (fake) rate in the particle identification (PID) of the experiment. For example, $K^-\to \pi^-X$ in-flight decay can be reconstructed as one track, where the hits from the daughter pions in the outer detector systems will dominate the reconstruction, hence misidentifying the mother kaon as a pion. In this paper, we refer to the primary track of the kink as mother and the secondary track of the kink as daughter. Furthermore, two-body kaon and pion decays can be used to improve the $dE/dx$ calibration, since the mother particle can be identified by a peak in the momentum distribution of the daughter particle in the mother's rest frame.

A well-known feature of the track-finding software is that it may accidentally reconstruct the trajectory of one single particle as two separate tracks, creating a pair of duplicates known as clone tracks. The point where these tracks intersect can mimic a kink vertex. Identifying and removing these clones improves tracking performance by reducing the double counting of a single particle. 

Finally, kinks are interesting for physics, e.g., muon in-flight decay reconstruction provides information about the muon polarization, which is important for the Michel parameters measurement in $\tau^-\to\mu^-\bar{\nu}_\mu\nu_\tau$ decays~\cite{Bodrov:2022mbd,Belle:2023udc,Belle:2023dyc}.

To our best knowledge, not many experiments have implemented an algorithm for searching for kinks; the most well-documented algorithm is provided by the NOMAD Collaboration~\cite{NOMAD:1997pcg} in Ref.~\cite{Astier:1999rs}. Unfortunately, the differences between the Belle II and the NOMAD experimental setups, in combination with the specific structure of the Belle II tracking software, including the \texttt{genfit2} track fitting package, does not allow for direct implementation of this algorithm. Thus, Belle~II needs its own, dedicated algorithm.

This paper is outlined as follows: In chapter~\ref{sec:belle2}, we give an overview of the Belle II experiment. In chapter~\ref{sec:kinksAtB2}, we discuss processes resulting in kinks in the Belle II detector, and in chapter~\ref{sec:kinkfinder}, we present an algorithm to reconstruct these. In chapter~\ref{sec:performance}, we present a performance study, and in chapter~\ref{sec:problems}, we identify problems and possible improvements. Finally, we present the conclusions in chapter~\ref{sec:conclusion}.

\section{Belle~II experiment}
\label{sec:belle2}
The Belle II experiment is located at the SuperKEKB asymmetric-energy $e^+e^-$ collider~\cite{Akai:2018mbz}. The Belle II detector consists of subsystems arranged cylindrically around the interaction point (IP)~\cite{Abe:2010gxa, Kou:2018nap}. Belle II uses right-handed orthogonal coordinates in which the $z$-axis is approximately collinear with the electron beam, the $x$-axis points horizontally away from the center of the accelerator rings, and the $y$-axis points vertically upward, perpendicular to the accelerator plane. In practice, cylindrical coordinates are used with the radial distance $r=\sqrt{x^2+y^2}$ and the azimuthal angle $\varphi$ in the $xy$-plane, measured from the $x$-axis. Hits used for reconstruction of charged-particle trajectories (tracks) are measured by a two-layer silicon-pixel detector (PXD) surrounded by a four-layer double-sided silicon-strip vertex detector (SVD) and a central 56-layer drift chamber (CDC). The latter two detectors also measure the energy loss from ionization. A quartz-based Cherenkov detector (TOP) identifies charged hadrons in the central region by measuring the time-of-propagation of the Cherenkov light emitted when the particles cross the quartz bars. An aerogel-based ring-imaging Cherenkov counter (ARICH) identifies charged hadrons in the forward region. An electromagnetic calorimeter (ECL) made of CsI(Tl) crystals measures photon and electron energies and directions. The above subdetectors are immersed in a 1.5 T axial magnetic field provided by a superconducting solenoid. A subdetector dedicated to identifying muons and $K_L^0$ mesons (KLM) is installed outside of the solenoid. 

The Kink Finder aims to reconstruct kinks inside the CDC. Therefore, a part of the mother track and the full daughter track are reconstructed based on CDC hits. The CDC inner (outer) radius is 17 (110) cm. It consists of 56 layers arranged in nine superlayers. The superlayers with wires aligned parallel to the solenoid magnetic field are called axial (A). To provide three-dimensional (3D) tracking, the stereo superlayers are used. They contain wires skewed by an angle between 66.8 and 74.1 mrad in the positive direction and $-58$ to $-78.6$ mrad in the negative direction, designated U and V, respectively. The resulting sequence of superlayers is A, U, A, V, A, U, A, V, A. 

For simulation, reconstruction, calibration, and physics analysis within Belle II, a central, modular C++/Python framework, called Belle II Analysis Software Framework (\texttt{basf2}), is used~\cite{Kuhr:2018lps, basf2-zenodo}. For the track reconstruction, Belle II employs a multi-step scheme implemented within \texttt{basf2}. The process begins with a CDC-only track-finding based on the Legendre algorithm~\cite{Alexopoulos:2008zza}, followed by attachment of SVD clusters with a combinatorial Kalman filter (CKF)~\cite{Mankel:1998uy, Mankel:2004yv, CMS:2014pgm, ATLAS:2017kyn}. The remaining SVD hits are used to reconstruct tracks with a standalone SVD track-finder using Sector Map and cellular automaton algorithms~\cite{Gardner:1970, Glazov:1993ur, Abt:2002he, Funke:2014dga}. The results are combined and extrapolated to the PXD with a second CKF. 
Prior to the extrapolation between different detectors and for the final combination, the tracks are fitted using the deterministic annealing filter (DAF) provided by the \texttt{genfit2} package~\cite{Bilka:2019ang,jojosito_2023_10358638}. More detailed information about the Belle~II track-finding algorithm is provided in Ref.~\cite{Bertacchi:2020eez}.

The benchmarking of the Kink Finder is carried out using $\tau^+\tau^-$ events, generated with \texttt{KKMC}~\cite{Jadach:1999vf} and \texttt{TAUOLA}~\cite{Jadach:1990mz} packages, and $B\bar{B}$ events, generated with \texttt{PYTHIA8}~\cite{Sjostrand:2014zea} and \texttt{EvtGen}~\cite{Lange:2001uf} packages. The full detector geometry, interactions of final state particles with the detector material, and the detector response are simulated using the \texttt{GEANT4} package~\cite{Agostinelli:2002hh}.

\section{Kinks at Belle~II}
\label{sec:kinksAtB2}

\begin{figure*}[!htb]
  \centering
  \includegraphics[width=0.75\linewidth]{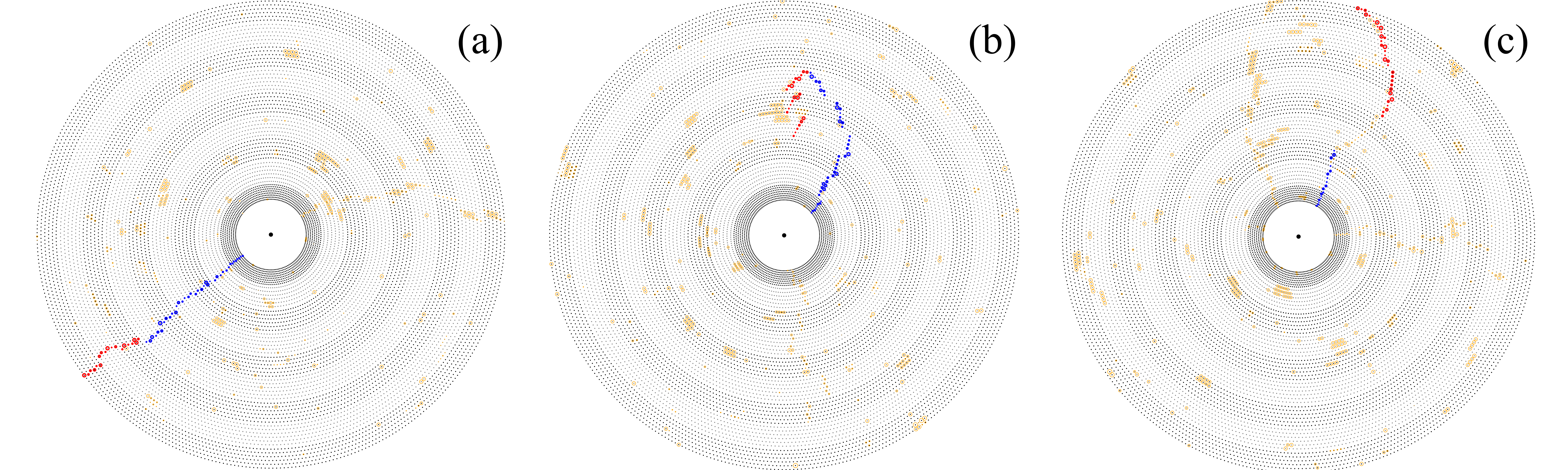}
\caption{CDC event display of three MC simulated events with kinks (a)--(c). Blue and red hits represent reconstructed mother and daughter tracks from the kink, respectively. All remaining hits in the CDC are yellow.}
\label{fig:1.1} 
\end{figure*}

Kinks at Belle~II can be classified into the following five categories based on the reconstruction of its components by the track-finding algorithm:
\begin{enumerate}
    \item both mother and daughter tracks are reconstructed separately;
    \item the mother track is reconstructed, while the daughter track is missing;
    \item the mother track is missing, while the daughter track is reconstructed;
    \item both mother and daughter tracks are missing;
    \item hits from mother and daughter tracks are combined into one reconstructed track. 
\end{enumerate}
The latter case is relevant to the PID fake rate, as mentioned in Sec.~\ref{sec:intro}. In around 10\% of in-flight decay events of kaons and pions from $\tau^-\to K^- \nu_\tau$ and $\tau^-\to \pi^- \nu_\tau$ decays, the daughter track cannot be reconstructed since it leaves too few hits inside the CDC. A similar fraction of in-flight decay events is obtained for average kaon and pion in $B\bar{B}$ events. Although some of the kink processes produce more than one charged daughter particle, it is usually possible to reconstruct and combine only one of them with the mother track, e.g., it is quite rare that more than one daughter pion is reconstructed from the \kcpi\ decay. Examples of MC simulated events with kinks inside the Belle~II CDC are shown in Fig.~\ref{fig:1.1}. The blue and red hits show the reconstructed mother and daughter tracks, respectively, while yellow hits represent all reconstructed hits inside the CDC, including the background.

We illustrate the experimental condition for kinks using MC samples containing $\tau^-\to K^- \nu_\tau$ and $\tau^-\to \pi^- \nu_\tau$ decays with the subsequent kaon and pion in-flight decays. For these in-flight decays, the reconstruction efficiency of the mother track depends on the process and is in general above 85\%. This is slightly lower compared to the reconstruction efficiency of ordinary tracks. Four layers of the SVD enable track reconstruction using SVD hits alone. This capability allows for the reconstruction of particles that decay at the inner edge of the CDC and recovers the efficiency for events where all CDC hits from the mother particle are accidentally assigned to the daughter track. The reconstruction efficiency of daughter tracks is much lower compared to that of mother tracks and varies from 14\% to 30\%, depending on the type of the decaying particle. Between 5\% and 20\% of the daughter tracks are reconstructed with the wrong charge since the Belle II track-finding algorithm, that also determines the charge, assumes that all tracks originate from the IP. Some hits from the mother track are often wrongly assigned to the daughter, and vice versa, which reduces the reconstruction efficiency and distorts the track parameter resolution. The probability of this to occur, as well as the number of hits, depend on the prominence of the kink: the less prominent it is, the higher the probability and number. 

There are around 40\% of kaon and 70\% of pion events containing in-flight decays, in which there is one reconstructed track arising from a combination of mother and daughter hits. Often, the hits attributed to either the mother or the daughter dominate in such combined track, allowing for its relation to the dominant particle at the MC level. Therefore, some of the combined tracks are included in the calculation of the reconstruction efficiencies above. Around 40\% of the combined tracks can neither be attributed to the mother nor to the daughter. This is because the ratio of the hit contributions is almost 50\%/50\%, whereas the Belle II definition requires at least 66\% of hits from a single particle to create a relation with it~\cite{Bertacchi:2020eez}.

\section{Kink Finder algorithm}
\label{sec:kinkfinder}

The Kink Finder algorithm developed at Belle~II addresses two general cases of kinks.
In the first case, both mother and daughter tracks are found individually by the Belle II track-finding algorithm. The Kink Finder aims at identifying these tracks, combining them, reconstructing the topology and the kinematics of the kink, and storing the related information required at the analysis level. In the second case, the hits from the mother and the daughter tracks are reconstructed as one combined track by the track-finding algorithm. The Kink Finder then aims at identifying such a track, splitting it into two separate tracks and then repeating the procedure for the first case. In the following sections, we describe the algorithm in detail.

\subsection{Reconstructed track pairs}\label{sub:general}
First, we describe the case with both mother and daughter tracks reconstructed individually. To optimize the execution time of the algorithm, we perform a preselection of tracks that are candidates to form a kink. The mother track candidate should start in the vicinity of the IP and, if it has CDC hits, end inside the CDC. The daughter track candidate should start (or end, if the assigned charge of the track is wrong) inside the CDC and its point of closest approach to the IP should be at a minimum distance.

Next, each mother candidate is combined with each daughter candidate, resulting in six different configurations as described below in the order of highest selection priority. The first three configurations assume good resolution of the track parameters of both candidates: 
\begin{enumerate}
    \item The starting point of the daughter track is close to the endpoint of the mother track in 3D [see Fig.~\ref{fig:1.1}(a)].
    \item The endpoint of the daughter track is close to the endpoint of the mother track in 3D [see Fig.~\ref{fig:1.1}(b)]. This configuration covers events with wrong charge assignment of the daughter track.
    \item The extrapolated daughter track is close to the endpoint of the mother track in 3D [see Fig.~\ref{fig:1.1}(c)]. This configuration covers events in which the daughter track intermediate hits are lost.
\end{enumerate}
The remaining three configurations are 2D equivalents of the former three: the only difference is in the comparison performed in the $r\varphi$-plane. The 2D approach serves the purpose to recover events in which the daughter track has a bad resolution in the $z$-coordinate, thereby improving the overall Kink Finder efficiency by about 30\%. The further processing of the algorithm depends on the selected configuration to increase the reconstruction efficiency and improve the kink spacial and momenta resolutions.

Several assisting algorithms are applied to improve the kink parameters resolution in the case of two separately reconstructed tracks. One example is hit reassignment between tracks and daughter track refitting with better initial values (seeds). We start with track pairs selected by the first three configurations (described above), all of which have good $z$-coordinate resolution for the daughter track. For these pairs, the processing begins with a vertex fit to find the location of the kink. The fitting procedure is described in Sec.~\ref{sub:vertexfit}. In the case of poor $z$-coordinate resolution of the daughter track, the track is refitted using the time and position information from the endpoint of the mother track. If the track charges are correctly reconstructed, and the start point and end point of the tracks are close to each other (normal ordering), the poor $z$-coordinate resolution of the daughter track may be due to wrong assignment of the first stereo superlayer in the CDC. We then try to refit the daughter track without these hits. Using the new fit result for the daughter, the vertex fit is performed.

If the vertex is located along one of the tracks for the normal ordering, the hit reassignment between them is done with a dedicated algorithm described in Sec.~\ref{sub:hitreassign}. We do not apply this algorithm to track pairs with the wrong charge of the daughter track, since the kink is quite sharp in this case and hence, wrong assignment of the hits inside CDC is unlikely. The algorithm cannot be used for the events in which the daughter track is reconstructed without hits from the inner superlayers.

In the case of wrong charge assignment of the daughter track, we flip the track and refit it with the time and position seed information from the mother track endpoint. If the fit fails, we proceed with the initial wrong charge combination, which can be corrected at the analysis level.

While the DAF fitting algorithm from the \texttt{genfit2} package is used as the default track fitter at Belle~II, we use the Reference Kalman Filter from the same package~\cite{Bilka:2019ang,jojosito_2023_10358638} for the fit of the flipped tracks and the refit of the daughter tracks with missing superlayer hits, as it shows a better performance.

The resulting kinks are filtered based on the radial position of the fitted kink location and the distance of the closest approach of the two tracks at the kink vertex. The first is required to be outside the VXD to suppress fake combinations of tracks from the IP and inelastic scattering of hadrons on the detector material.

To suppress pairs of clones mimicking a kink, a track is formed and fitted by combining hits from both tracks. The following fit information, sufficient to suppress clones at the analysis level, is saved in the form of a bitmask. Each bit in this mask stores a Boolean value (0 or 1) representing the result of a specific test:
\begin{enumerate}
    \item \textbf{Success of the combined track fit:} A Boolean flag indicating whether the track fit converged successfully (1) or failed (0).
    \item \textbf{Fit quality:} The remaining bits store the results of comparisons of different fit quality markers. For each quantity below, the bit is set to 1 if the condition is true and 0 if false:
    \begin{itemize}
        \item The number of degrees of freedom (n.d.f.) of the combined track is greater than the initial mother track's n.d.f.
        \item The n.d.f. of the combined track is greater than the initial daughter track's n.d.f.
        \item The p-value of the fitted combined track~\cite{Bertacchi:2020eez} is greater than the initial mother track's p-value.
        \item The p-value of the fitted combined track is greater than the initial daughter track's p-value.
        \item The p-value of the fitted combined track is greater than a predefined cut value of $10^{-7}$ (optimized for Belle II).
    \end{itemize}
\end{enumerate}

\subsubsection{Vertex fit}\label{sub:vertexfit}

The fit of the vertex position is performed using a kinematic fitter algorithm implemented in the Belle~II Software. It is based on the least squares method and described in Ref.~\cite{Tanaka:2001ae}, Appendix D. The \texttt{genfit2} package also provides an algorithm for the vertex fitting, \texttt{GFRave}~\cite{Bilka:2019ang,jojosito_2023_10358638,Waltenberger:2011zz}, which is applied at Belle~II for the V0 ($K^0_S$, $\Lambda$, and externally converted $\gamma$) vertex reconstruction. However, we do not use it in the Kink Finder since it does not have an option for setting an initial value. Without such an option, the vertex fit fails for a noticeable amount of kinks.

As initial vertex position for the fit, we use the point of the closest approach (POCA) of the mother track, calculated at its last assigned hit. This hit is the farthest from the IP along the track's trajectory. In the further fits, the previous vertex fit result is used as a seed. The vertex fit is applied to the helices of both tracks determined at the seed vertex taking energy loss and the complete magnetic field map into account. After the fit is converged, both tracks are extrapolated to the vertex. If the vertex is located along one of the tracks, the hit reassignment is performed.

\subsubsection{Hit reassignment algorithm} \label{sub:hitreassign}

This algorithm is responsible for the reassignment of hits between two tracks and serves to improve the kink spacial and momenta resolutions and reconstruction efficiency. It is performed in an iterative procedure, where each iteration contain the following steps:
\begin{enumerate}
    \item {\bf Identify the ``kink'' hit:} Find the ``kink'' hit on the mother (or daughter) track that is closest to the fitted kink location. The choice between the mother or daughter track is determined by which track's trajectory the kink position lies on. All hits after ``kink'' hit on the mother track, or before it on the daughter track, are flagged for reassignment to the complementary track.
    \item {\bf Reassign Hits:} Update the hit assignments for both tracks based on the ``kink'' hit identified in the previous step.
    \item {\bf Refit Tracks:} Fit the tracks with new hit sequences. The fit must converge successfully. The quality of the new hit configuration is evaluated by the ratio of the combined $\chi^2$ to the combined n.d.f. defined as
    \begin{equation} \label{eq:5.1}
        R_\text{comb}=\dfrac{\chi^2_\text{comb}}{\text{n.d.f.}_\text{comb}}=\dfrac{\chi^2_\text{m} + \chi^2_\text{d}}{\text{n.d.f.}_\text{m} + \text{n.d.f.}_\text{d}},
    \end{equation}
    where the indices ``m'' and ``d'' correspond to the mother and daughter tracks, respectively. The new $R_\text{comb}$ value should be smaller than the one calculated for the initial pair.
     \item  {\bf Handle Failures:} If the fit in Step 3 fails or does not yield a better $R_\text{comb}$, shift the ``kink'' hit one position closer to the second track and return to Step 2. If this rollback occurs more than five times in a single iteration, the hit reassignment is considered a failure. This step prevents the influence from bad hits or other possible problems with the fit.
     \item {\bf Update Vertex:} Perform a vertex fit. If no further hit reassignment is required, the iterative procedure is terminated.
\end{enumerate}
The number of total iterations is limited to three for the Belle~II to reduce the usage of computer resources.

\subsection{Combined tracks}\label{sub:splitting}
In the case when mother and daughter hits are combined into one track, a track candidate is selected based on the following criteria before further splitting, filtering, and storing: 
\begin{enumerate}
    \item The p-value of the track fit~\cite{Bertacchi:2020eez} must be lower than the threshold, which is set several orders of magnitude larger than the loose threshold used in general track fitting. The value used in the Kink Finder is optimized for the experimental conditions to suppress the splitting of one true track into two false tracks.
    \item The number of fitted CDC hits in the track must be less than the threshold also optimized for the experimental conditions to suppress the splitting of one true track into two false tracks.
    \item The track distance of the closest approach to the IP must be small. However, the track itself is not required to start in the IP region to allow for cases where the CDC track segment has poor resolution, causing the extrapolation into the VXD (PXD and SVD together) to fail.
\end{enumerate}
Then, the combined track candidates are split using the algorithm that creates a pair of mother and daughter tracks using the hits of the initial track.

First, the initial splitting is performed at three points along the track, identified by the following ratios of the mother-to-daughter fractions of the initial number of hits: $80\%/20\%$, $50\%/50\%$, and $20\%/80\%$. Then, three created pairs of tracks are fitted. After each fit of the track pair in the described procedure, the value of 
$|R_\text{comb}-1|$ is calculated based on the fit and compared to the value of $|\chi^2_\text{in}/\text{n.d.f.}_\text{in}-1|$ for the initial track. If the former value is bigger, the obtained track pair is considered as ``bad''. 

After this preparation, the actual search of the splitting point is performed in five iterations, which is usually enough for convergence. Each iteration is based on the following binary search:
\begin{enumerate}
    \item Choose the first or the third splitting point (edge) with the smallest $|R_\text{comb}-1|$ value. If both track pairs are labeled as ``bad'', then check the value for the middle point. If it is also ``bad'', the splitting procedure fails; otherwise, the middle point is taken as a final splitting point.
    \item Find a new middle point between the selected edge and the current middle point. Split the track in a new point and fit the resulting track pair.
\end{enumerate}
The procedure is repeated until convergence or reaching the limit. The best track splitting point is defined as the one giving the smallest $|R_\text{comb}-1|$ value. The resulting track pair is then used to find the kink location using the same fitting procedure as before (see Sec.~\ref{sub:vertexfit}). The filtering of the result is based only on the radius of the fitted kink vertex: it should be outside the VXD, as in the case of track pairs.

\section{Performance}
\label{sec:performance}
The Kink Finder performance for in-flight decays is studied using MC samples of the $\tau^-\to K^- \nu_\tau$ and $\tau^-\to \pi^- \nu_\tau$ events with a total number of $8\times10^4$ kaon and $13\times10^4$ pion in-flight decays inside the CDC. The performance for general cases of kinks is studied using MC samples of generic $\tau$-pair and $B\bar{B}$ decays with $10^5$ events in each sample. In both studies, we account for beam-induced background, which consists of spurious detector hits and particles produced in secondary interactions with the detector and support material outside the IP region~\cite{Natochii:2022vcs}. The level of this background strongly depends on the SuperKEKB luminosity and its conditions. In our studies, we use a configuration that is considered typical for current Belle II operations. Although the Kink Finder has been included in the Belle~II tracking software, it has not been applied for data processing yet, and therefore, there is no performance study using data available at the time of writing.

\subsection{Reconstructed track pairs}
\label{sub:performance_2}

In this part, we provide the results of the performance study for the Kink Finder applied to the events with both daughter and mother tracks being found by the track-finding algorithm. We compare the spacial and momenta resolutions obtained by the Kink Finder to those from a default method, which uses only a vertex fit of two tracks at the analysis level.

\begin{figure}[htb]
  \centering
  \includegraphics[width=1\linewidth]{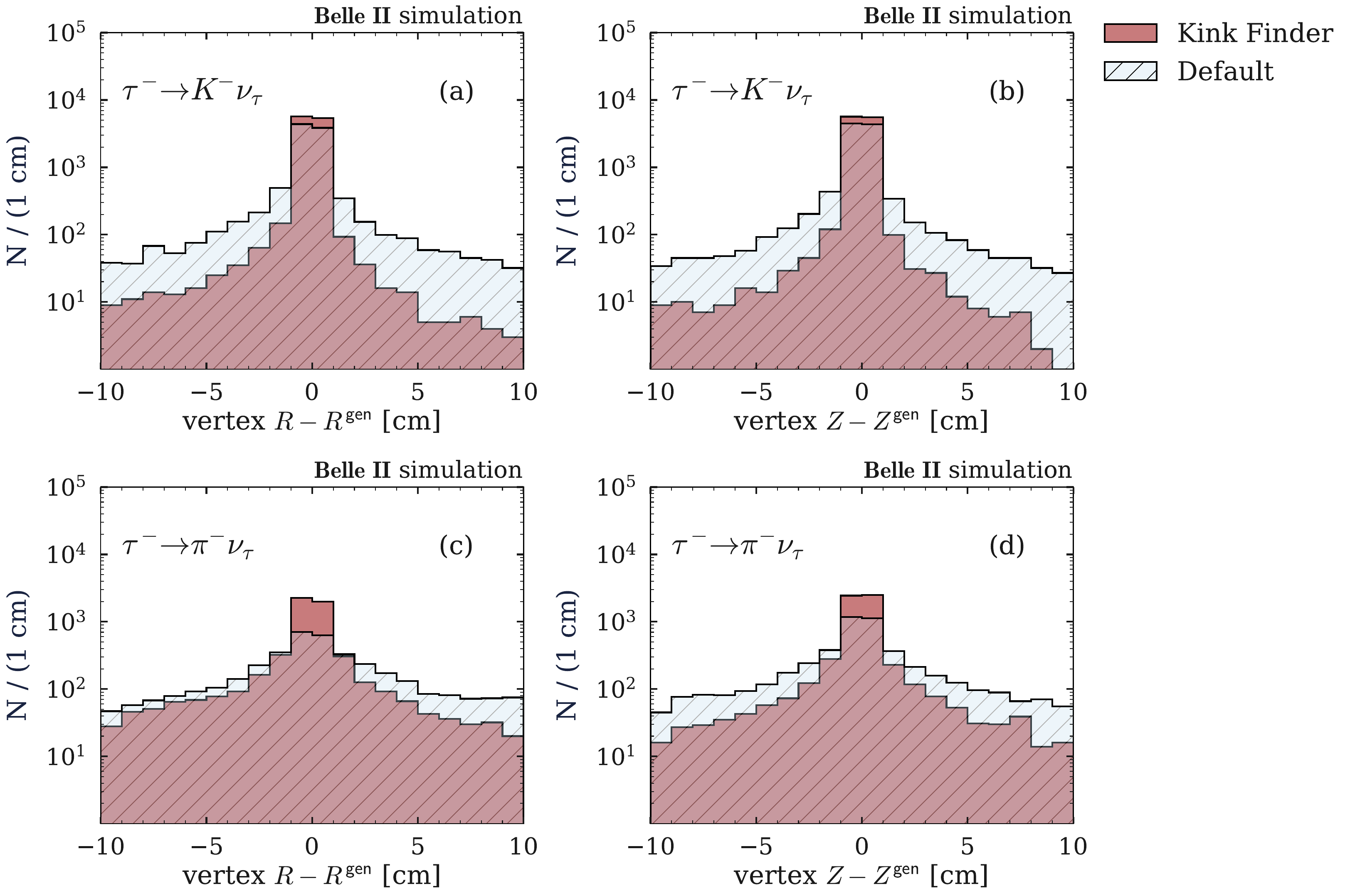}
\caption{Resolution of the $r$- and $z$-components of the kink vertex position for in-flight decays. Panel (a,~b) are obtained with the $\tau^-\to K^-\nu_\tau$ sample and panel (c,~d) with the $\tau^-\to \pi^-\nu_\tau$ sample.}
\label{fig:8.1} 
\end{figure}

The main parameters of the in-flight-decay kink are the decay vertex position $(x,\,y,\,z)$, the corresponding momenta of the mother and the daughter particles, and the momentum of the daughter particle in the mother particle rest frame. Figure~\ref{fig:8.1} shows the kink vertex $r$- and $z$-component resolutions for kaon and pion decays. The overall resolution of the vertex position, defined as the spread of the distribution of the distance from the real decay vertex to the reconstructed one, is 0.9 cm for kaon decays and 2.4 cm for pion decays. Approximately 93\% and 62\% of events are contained within a 1 cm region in the $rz$-plane, respectively. A better resolution for kaons is attributed to the larger energy release in the decay, leading to a more prominent kink that is easier to fit.

\begin{figure}[!htb]
  \centering
  \includegraphics[width=1\linewidth]{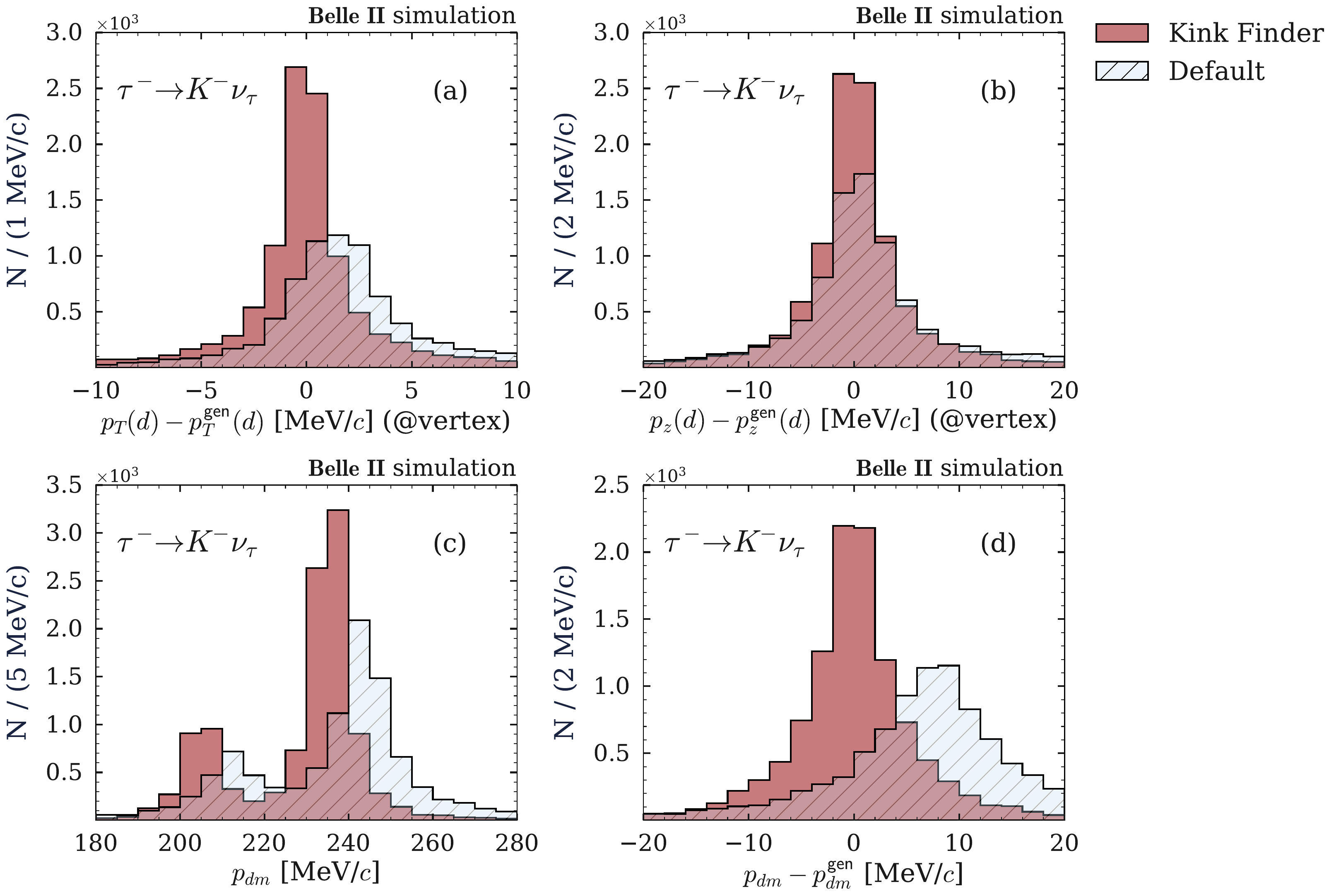}
\caption{Resolution of the daughter momentum in the laboratory frame for transverse component (a) and $z$-component (b); distribution (c) and resolution (d) of the daughter momentum in the mother rest frame. The plots are obtained with the $\tau^-\to K^-\nu_\tau$ sample. The two peaks in (c) correspond to the $K^-\to\pi^-\pi^0$ (left) and $K^-\to\mu^-\bar{\nu}_\mu$ (right) decays.}
\label{fig:8.3} 
\end{figure}

\begin{figure}[!htb]
  \centering
  \includegraphics[width=1\linewidth]{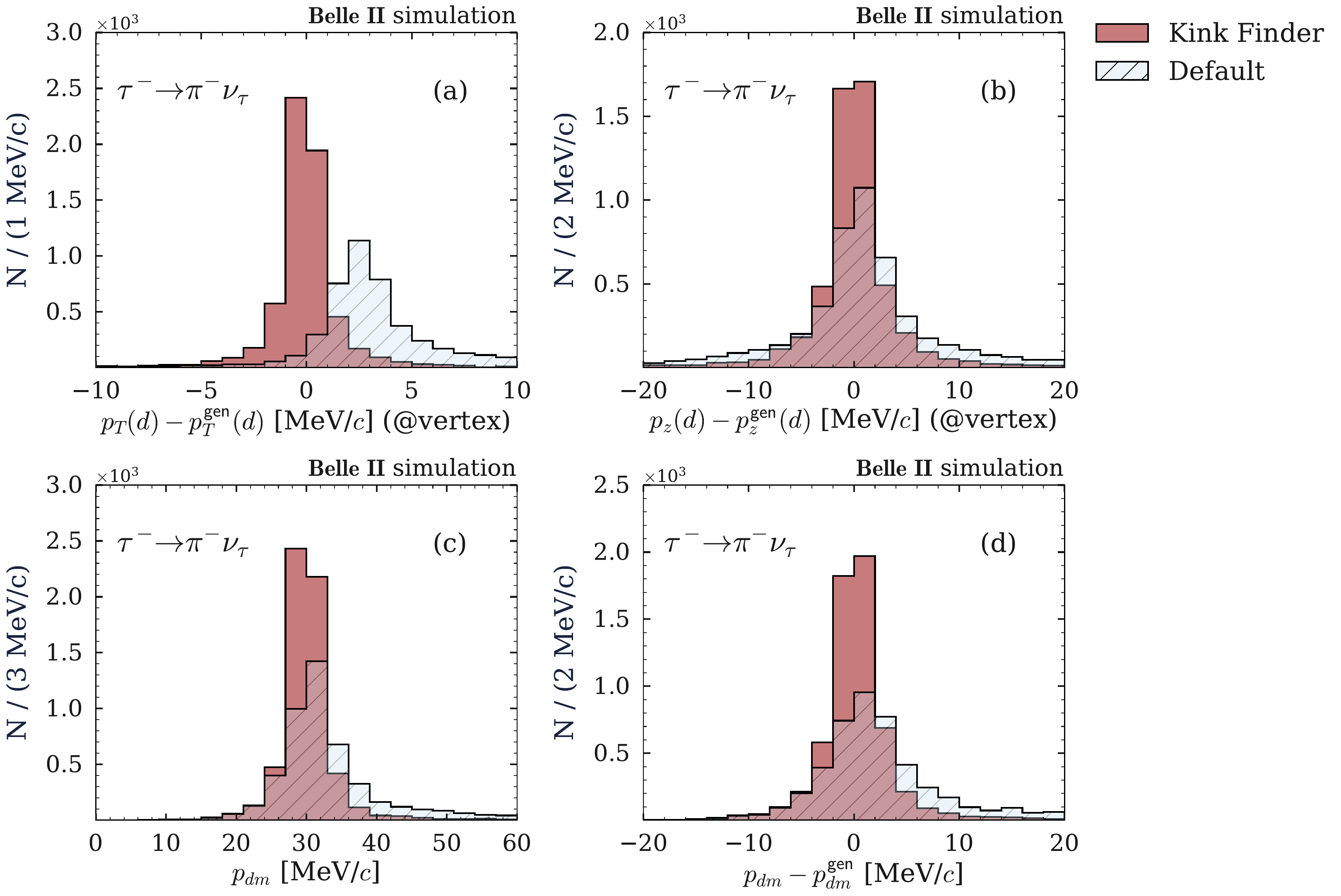}
\caption{Resolution of the daughter momentum in the laboratory frame for transverse component (a) and $z$-component (b); distribution (c) and resolution (d) of the daughter momentum in the mother rest frame. The plots are obtained with the $\tau^-\to \pi^-\nu_\tau$ sample.}
\label{fig:8.4} 
\end{figure}

The daughter momentum resolutions in the laboratory frame ($p^\text{lab}$) and in the mother particle rest frame ($p_{dm}$) for kaon and pion two-body decays are shown in Fig.~\ref{fig:8.3} and~\ref{fig:8.4}, respectively.
For both samples, a significant improvement is obtained with the application of the Kink Finder. Better resolutions for the daughter track parameters are achieved with the hit reassignment and improved estimation of the track seed time used for the track extrapolation based on the information from the last hit of the mother track. For the daughter particle momentum in the mother particle rest frame, there is an additional positive effect from the better accuracy of the decay vertex measurement, improving the determination of the angle between daughter and mother momenta. Finally, in the default case, there is a bias of $2\mevc$ ($3\mevc$) in the transverse momentum of the daughter particle and $9\mevc$ ($2\mevc$) in the $p_{dm}$ momentum for kaon (pion) decays caused by energy losses corrections. This bias is completely (up to statistical uncertainty) eliminated by the Kink Finder. Figure~\ref{fig:8.3}(c) shows two distinct peaks originating from $K^-\to\pi^-\pi^0$ and $K^-\to\mu^-\bar{\nu}_\mu$ decays. Similarly, a single peak from $\pi^-\to\mu^-\bar{\nu}_\mu$ decay is clearly identified in Fig.~\ref{fig:8.4}(c).

Using generic $\tau$-pair and $B\bar{B}$ samples, we estimate the efficiency of the Kink Finder for the reconstruction of an in-flight decay; the result is shown in Table~\ref{tab:8.1}. While the reconstruction efficiency for kaon decays is around 80\%, it is only around 50--65\% for pion decays, depending on the process where the pion is produced. This difference is caused by the low energy release in the \pimunu\ decay. The higher track multiplicity of the $B\bar{B}$ events reduces the Kink Finder efficiency. The main reason for the
observed inefficiency is tracks with poor $z$-resolution, which cannot be recovered by 2D selection described in Sec.~\ref{sub:general}. 
\begin{table}[t]
\renewcommand{\arraystretch}{1.15}
\centering
\caption{Total number of in-flight decays inside the CDC for samples of $10^5$ $\tau$-pair and $B\bar{B}$ events ($N^\text{MC}$), number of in-flight decays with both mother and daughter tracks reconstructed by the track-finder ($N^\text{MC}_\text{reco}$), and number of in-flight decays found by the Kink Finder ($N_\text{kinks}$).} \label{tab:8.1}
\begin{tabular}
 {@{\hspace{0.25cm}}c@{\hspace{0.25cm}} @{\hspace{0.25cm}}c@{\hspace{0.25cm}}  @{\hspace{0.25cm}}c@{\hspace{0.25cm}}
 @{\hspace{0.25cm}}c@{\hspace{0.25cm}}
 @{\hspace{0.25cm}}c@{\hspace{0.25cm}}}
\hline \hline
& $N^\text{MC}$ & $N^\text{MC}_\text{reco}$ & $N_\text{kinks}$ & $N_\text{kinks}/N^\text{MC}_\text{reco}$ \\
\hline
\multicolumn{5}{c}{$\tau$-pair sample}\\
All decays & 6212 & 666 & 470 & 71\%\\
Pion decays & 5803 & 581 & 390 & 67\% \\
Kaon decays & 365 & 85 & 80 & 94\% \\
\multicolumn{5}{c}{$B\bar{B}$ sample}\\
All decays & 97144 & 10223 & 6429 & 63\%\\
Pion decays & 64754 & 5173 & 2617 & 51\% \\
Kaon decays & 31599 & 5046 & 3809 & 75\% \\
\hline \hline
\end{tabular}
\end{table}

In Table~\ref{tab:8.1}, we also provide the overall number of events with in-flight decays inside CDC at Belle~II: around 6\% of $\tau$-pair events have secondary particles decaying in flight. This is in contrast to the $B\bar{B}$ sample, where in-flight decays occur in almost all events. In both cases, the fraction of events with in-flight decays with a reconstructed mother and daughter, is around 11\%.

The total amount of two-track kinks found in $10^5$ generic $\tau$-pair and $B\bar{B}$ events is $2.4\times10^3$ and $1.7\times10^4$, respectively. The number of track pairs with correct mother-daughter relation makes up more than half of the reconstructed kinks ($61\%$ and $53\%$ for $\tau$ and $B\bar{B}$ pairs, respectively). Among them, in-flight decays constitute $32\%$ for $\tau$ pairs and $74\%$ for $B\bar{B}$ pairs, while the remaining kinks are from inelastic hadron scattering. The majority of events with inelastic hadron scattering occur in the VXD material and can be easily suppressed at the analysis level. An additional $27\%$ ($31\%$) of kinks in the $\tau$-pair ($B\bar{B}$) sample are combinations of a track from IP with its clone. In both samples, around $2\%$ ($9\%$) of the kinks have a fake track as a mother (daughter). Finally, the fraction of false combinations of two unrelated tracks is $2\%$ for the $\tau$-pair sample and $ 6\%$ for the $B\bar{B}$ sample.

\begin{figure*}[!htb]
  \centering
  \includegraphics[width=0.65\linewidth]{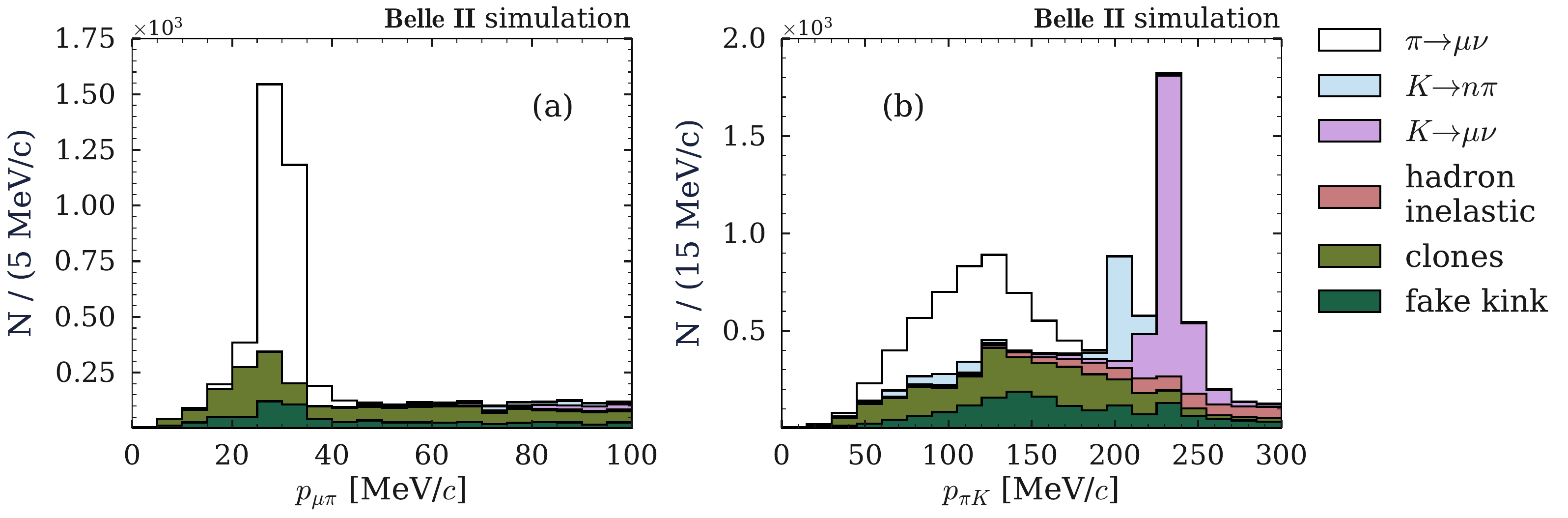}
\caption{Daughter particle momentum in the mother's rest frame with muon and pion mass hypotheses (a) and pion and kaon mass hypotheses (b). The various sources of kinks in $B\bar{B}$ sample are shown as stacked histograms.}
\label{fig:8.5} 
\end{figure*}

In Fig.~\ref{fig:8.5}, we illustrate the composition of the reconstructed kinks in the $B\bar{B}$ sample by providing the distributions of the daughter particle momentum in the mother particle rest frame for two pairs of different mass hypotheses: muon (pion) and pion (kaon) for daughter (mother) particles, corresponding to \pimunu\ and \kpipi\ decays. 
Here, we have further suppressed the contribution from clones, using the output of the combined fit described in Sec.~\ref{sub:general}. Narrow peaks from two-body pion and kaon decays are clearly discerned. These distributions can be compared with the corresponding ones from Belle, obtained without the Kink Finder~\cite{Belle:2023dyc}. A significant improvement in the resolution is observed, which creates better conditions for physics studies, e.g., the Michel parameter $\xi^\prime$ measurement.

The combined fit of two tracks allows for a 60\% suppression of the clones with a simple veto on the saved flag, introduced in Sec.~\ref{sub:general}. Additional information like the difference of the daughter and mother particle momenta can be exploited to further suppress clones at the analysis level.

\subsection{Combined tracks}
\label{sub:performance_1}

\begin{figure}[htb]
  \centering
  \includegraphics[width=1\linewidth]{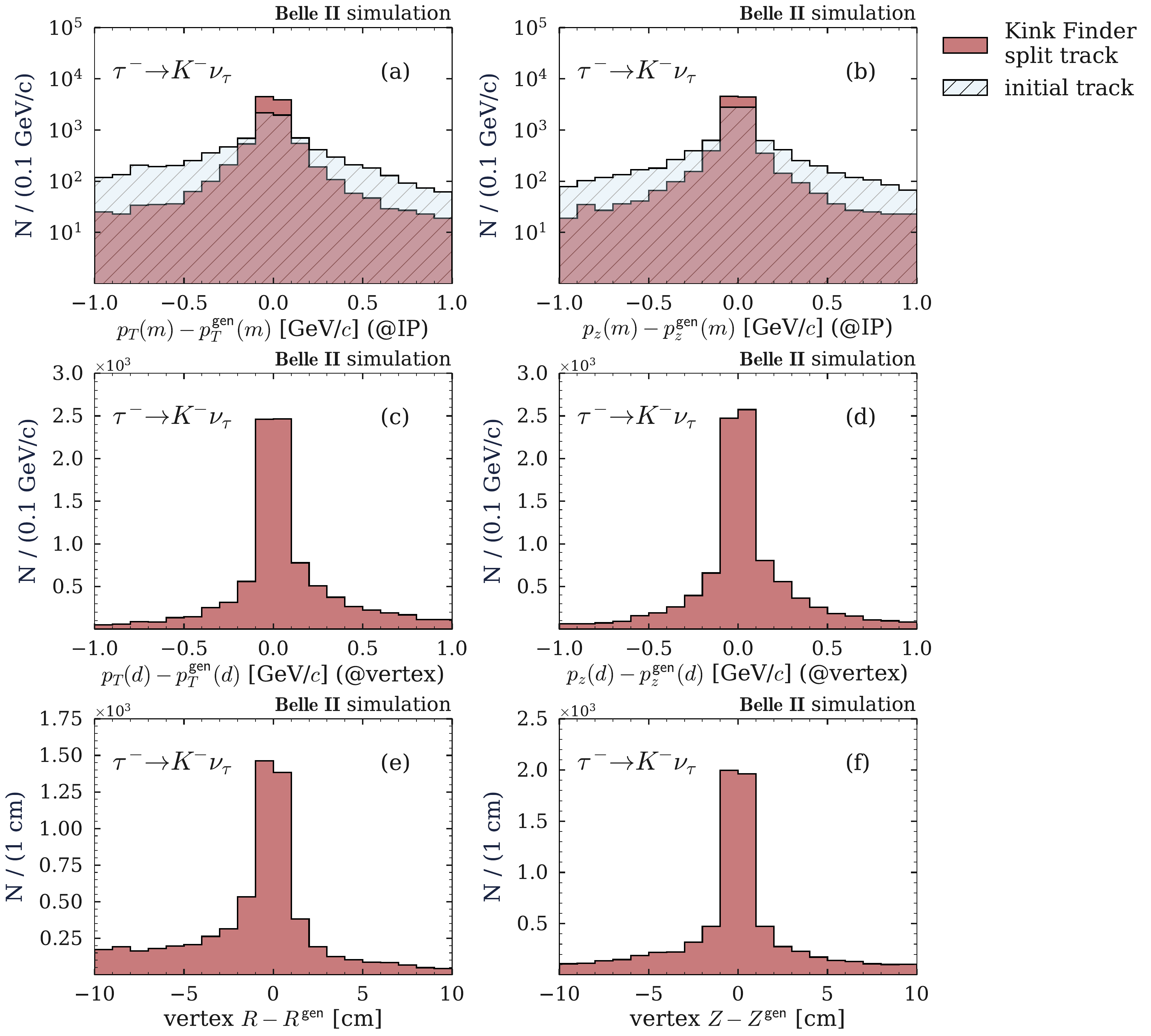}
\caption{Transverse and longitudinal momentum resolution of mother particle, panel (a,~b), and daughter particle, panel (c,~d). Resolution of the $r$- and $z$-components of the vertex position, panel (e,~f). The plots are obtained with the $\tau^-\to K^-\nu_\tau$ sample.}
\label{fig:8.8} 
\end{figure}

The efficiency of the Kink Finder algorithm for combined tracks from pion decays in the $\tau^-\to \pi^-\nu_\tau$ sample and kaon decays in the $\tau^-\to K^-\nu_\tau$ sample is found to be around 53\% and 39\%, respectively. The Kink Finder splitting algorithm improves the momentum resolution of the mother track compared to the initial track,\footnote{As the initial track comes from IP, we compared it to the mother track of the kink.} as illustrated in Fig.~\ref{fig:8.8}(a) and (b) for kaon decays. The resolutions of the daughter track momentum and the decay vertex for kaon decays are shown in Fig.~\ref{fig:8.8}(c)--(f). The splitting algorithm correctly reconstructs the central value, although with a significantly larger spread compared to the kinks reconstructed from track pairs. The result for pion decays is the same.
Despite the worse resolution, a larger yield of this kind of kinks can benefit the analysis of the in-flight decays.

In addition, the mere fact that a track has been split by the Kink Finder is sufficient to reduce the PID fake rate. The fake rate of the standard Belle~II Neural-Network-based binary PID~\cite{Belle-II:2025tpe} is studied for kaons and pions using the $B\bar{B}$ sample. The exclusion of particles found by the Kink Finder improves the overall kaon and pion fake rates by 1\% and 5\%, respectively; however, the effect depends on the transverse momentum $p_T$. For pions, the main improvement is in the low-$p_{T}$ region $p_T<1.2\,\text{GeV}/c$, with a maximum of 10\% for $p_T<0.6\,\text{GeV}/c$. For kaons, the improvement is negligible for the interval $0.6\,\text{GeV}/c<p_T<1.2\,\text{GeV}/c$, while for the remaining region, it is around 2--3\%. We have also studied the potential maximal reduction of the fake rate if all in-flight decays, reconstructed as one combined track, are found by the Kink Finder. For pions, the largest effect is again obtained for low $p_T$ particles with a maximum of 15\% for $p_T<0.6\,\text{GeV}/c$. For kaons, the expected improvement is $\approx3\%$ in the same region as for the veto on tracks found by the Kink Finder. Thus, we conclude that the in-flight decays in general have quite a small effect on the PID fake rate.

The ratio of the false splitting to the correct one is obtained using generic $\tau$-pair and $B\bar{B}$ samples to be 4.2 and 2.1, respectively. The observed large contribution to the false splitting is caused by a significant amount of ordinary tracks with a bad p-value used for the preselection of the combined track candidates. At the analysis level, this ratio can be suppressed using machine learning (ML) algorithms trained on the information about initial and resulting tracks and kink characteristics.
 
\section{Problems and further improvement}
\label{sec:problems}

In this section, we discuss the existing problems of the developed Kink Finder algorithm and ideas on possible solutions. 

The current Kink Finder algorithm efficiently reconstructs most in-flight-decay kinks where both mother and daughter tracks are found by the track-finding algorithm. Further gains can be achieved by recovering events where the track-finding algorithm omitted an entire superlayer. Adding the missing stereo hits will improve the daughter track's $z$-coordinate resolution. This optimization is planned for future implementation.

The track-splitting efficiency is currently hampered by ordinary tracks misidentified as kink events. While the current strict preselection reduces the number of such events, it also has the disadvantage that it suppresses genuine kinks. Machine learning approaches will be explored with the aim to better distinguish the kink signatures.

While the mother track is reconstructed in the majority of kink events, daughter tracks have relatively low efficiency. Studies show that the application of the Belle~II segment finder and the following local track-finder~\cite{Bertacchi:2020eez} can increase the daughter track reconstruction efficiency by up to a factor of two. However, at present, the local track-finder is switched off due to high fake rate and clone rate. An upgrade, including the re-enabling of this feature, is under discussion. 

The Graph Neural Network (GNN) track-finding, recently developed and studied at Belle~II~\cite{Reuter:2024kja}, has shown an impressive improvement in the finding efficiency of the secondary tracks. Presently, this algorithm is being implemented in \texttt{basf2}. It is also considered to integrate kinks in the GNN training to increase the Kink Finder efficiency.

The execution of the Kink Finder is relatively slow, taking 13 ms and 111 ms per call for the $\tau$-pair and $B\bar{B}$ samples, respectively. This performance is up to two times slower than the Belle II algorithm for finding V0 vertices when normalized by the number of candidates. The primary reason is the computationally intensive iterative track fitting. To accelerate hit reassignment and track-splitting and improve the vertex convergence, a fitter modification is proposed that caches the fit result after each new hit is added during Kalman filtering. This cache enables a single forward-backward pass required for the procedure convergence. This approach has to be studied in the near future.

Finally, the precision of the kink parameters can be enhanced by using the geometric constraint of coupled helices at the vertex. A Kalman filter, applied to all hits using the information about the continuity of the kink (inspired by Ref.~\cite{Astier:1999rs}), would improve vertex and momentum resolutions. Implementation of two features described above within the \texttt{genfit2} package requires further development and is left for future updates of the Kink Finder.

\section{Conclusion}
\label{sec:conclusion}

In conclusion, the first version of the Kink Finder has been developed and implemented in the Belle~II tracking software. We have provided a detailed description of the algorithms and conducted a performance study using MC events. The overall efficiency of the reconstruction of the in-flight-decay kinks is obtained to be around $40\%$ for both kaon and pion decays. The resolution of the kink track parameters is improved compared to the default case, where the kink can be reconstructed at the analysis level without the Kink Finder. The obtained efficiency and resolutions can significantly improve the precision of the Michel parameter $\xi^\prime$ measurement compared to the current result by Belle. The number of cloned tracks identified by Kink Finder is reduced by a factor of two.
The PID fake rate is reduced for low-energy pions while the corresponding impact in kaons is marginal. We have identified several ways to further improve the Kink Finder and will implement these in the near future.

Regarding the application of the Kink Finder for the data analysis, the Belle~II data has not been reprocessed with the Kink Finder yet due to the Belle~II software release schedule. According to the current plans, it may take at least 1 or 2 years from now until the data with the applied Kink Finder is available for analysts. Even more time will be required before the first analyses of kinks are complete.

\section*{Acknowledgements}
The work is supported by the National Natural Science Foundation of China (NSFC) under Contracts Nos. 12335004 and 12575092.
The article was prepared within the framework of the project ``Mirror Laboratories'' HSE University.

This work, regarding the Belle II detector, which was built and commissioned prior to March 2019, was supported by
Horizon 2020 Marie Sklodowska-Curie RISE project JENNIFER2 Grant Agreement No.~822070 (European grants);
BMFTR, DFG, HGF, MPG, and AvH Foundation (Germany);
Emmy-Noether Grant No. 526218088;
Department of Atomic Energy under Project Identification No.~RTI 4002,
Department of Science and Technology,
and
UPES SEED funding programs
No.~UPES/R\&D-SEED-INFRA/17052023/01 and
No.~UPES/R\&D-SOE/20062022/06 (India);
Istituto Nazionale di Fisica Nucleare and the Research Grants BELLE2,
the ICSC – Centro Nazionale di Ricerca in High Performance Computing, Big Data and Quantum Computing, funded by European Union – NextGenerationEU;
The Knut and Alice Wallenberg Foundation (Sweden), Contracts No.~2021.0174, No.~2021.0299, and No.~2023.0315;
and the U.S. Department of Energy and Research Awards
No.~DE-SC0019230, 
No.~DE-SC0023470. 

These acknowledgements are not to be interpreted as an endorsement of any statement made
by any of our institutes, funding agencies, governments, or their representatives.

\bibliographystyle{elsarticle-num} 
\bibliography{references}

\end{document}